\documentclass{elsart}

    \usepackage{latexsym,bm,amsmath,amssymb,amsfonts}
    \usepackage{epsfig,graphics,graphicx}
    \usepackage{makeidx}
    \usepackage{citesort}

    \renewcommand{\theequation}{\thesection.\arabic{equation}}

\long\def\comment#1{ }

\newcommand{\beq}{\begin{eqnarray}}
\newcommand{\eeq}{\end{eqnarray}}
\newcommand{\be}{\vspace{-.4cm}\begin{eqnarray}}
\newcommand{\ee}{\vspace{-.5cm}\end{eqnarray}}

\newcommand{\cal}{\mathcal} 

\newcommand{\BQ}{\begin{equation}}
\newcommand{\EQ}{\end{equation}}
\newcommand{\BQA}{\begin{eqnarray}}
\newcommand{\EQA}{\end{eqnarray}}

\def\simge{\mathrel{%
   \rlap{\raise 0.511ex \hbox{$>$}}{\lower 0.511ex \hbox{$\sim$}}}}
\def\simle{\mathrel{
   \rlap{\raise 0.511ex \hbox{$<$}}{\lower 0.511ex \hbox{$\sim$}}}}

\begin{document}

\begin{flushright}
~\vspace{-1.25cm}\\
 SACLAY-T07/014\\
 CU--TP--1176
\end{flushright}
\vspace{2.cm}

\begin{frontmatter}

\parbox[]{16.0cm}{ \begin{center}
\title{Correlation of small--x gluons  \\ in impact parameter space}

\author{Y.~Hatta$^{\rm a}$ and A.~H.~Mueller$^{\rm b}$ }

\address{$^{\rm a}$ Service de Physique Th\'eorique,
     CEA/Saclay,
      \\ 91191 Gif-sur-Yvette Cedex,
              France }

\address{$^{\rm b}$ Department of Physics, Columbia University
\\ New York, NY 10027, USA }


\begin{abstract}
In the framework of the QCD dipole model at high energy, we
present an analytic evaluation of the dipole pair density in two
limits in which the parent dipole is much larger/smaller than the
distance between the two child dipoles. Due to conformal symmetry,
the two limits give an identical result. The power--law
correlation between dipoles explicitly breaks the factorization of
 target--averaged
 scattering amplitudes.
\end{abstract}
\end{center}}

\end{frontmatter}

\section{Introduction}
A hadron in the infinite momentum frame is a complicated system of
small--x gluons. While the energy evolution of the average
  gluon number
 can be described by the
  Balitsky--Fadin--Kuraev--Lipatov (BFKL)
 equation \cite{bfkl}, the wavefunction of a hadron
  contains more information than just the average number.
 For example, the fluctuation of the gluon number plays a crucial role
  in the evolution of scattering amplitudes towards the unitarity limit, and has
 recently attracted considerable
  interest
  \cite{salam,IM,imm,shoshi,edmond,golec,diff,liouville}
    in the context of saturation physics \cite{sat,qiu,larry}.

\begin{figure}
\begin{center}
\centerline{\epsfig{file=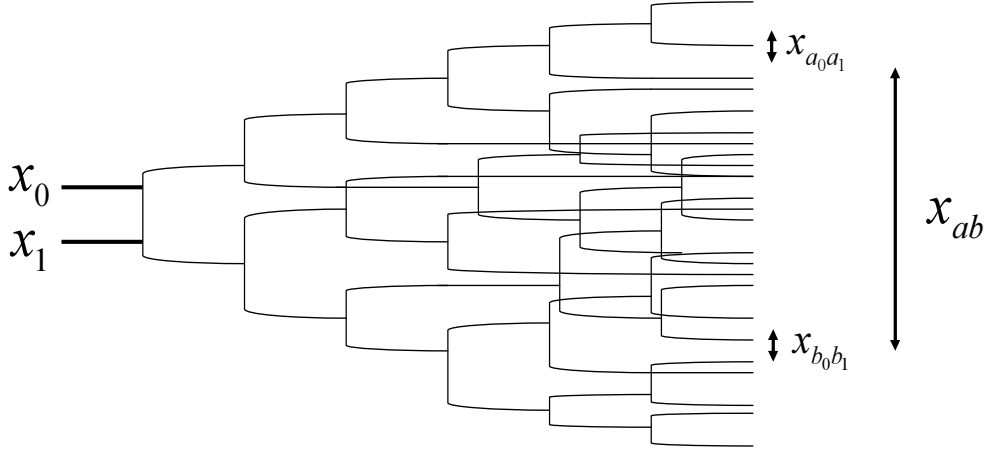,height=6.cm,width=13.cm}}
\caption{\sl
 A cascade of dipole splitting. Thin lines represent child dipoles
 produced by the parent dipole $(x_0,x_1)$.
 \label{fig1}}
\end{center}
\end{figure}

 Another important characteristic of the hadron
  wavefunction is the correlation of gluons in the
  impact parameter space. In the dilute, non--saturated regime,
  soft gluons are necessarily correlated because they originate from
a common ancestor via gluon splitting. The process can be most
easily described in the QCD dipole model formulated in the large
$N_c$ approximation \cite{al1,patel,al}. In this approach, the
evolution of the `parent' dipole (a quark--antiquark pair)
proceeds via dipole splitting with certain probability computed in
perturbation theory (Fig.~\ref{fig1}). Since the probability
depends nontrivially on transverse coordinates,  `child' dipoles
will be distributed in the transverse plane with characteristic
correlations between them. Although this dynamics is built--in in
the numerical Monte--Carlo simulation of this model
\cite{salam,av}, so far there have been only few analytical
insights \cite{patel,al,1,bialas}. (See, also, \cite{tre}.) In
this paper we evaluate the \emph{dipole pair density}  in certain
limits and find the power--law correlation between dipoles at
large distances with the power determined by the conformal weights
of the BFKL eigenfunction \cite{lev}.  As an immediate consequence
of our result, we shall show in Eqs.~(\ref{ka}) and (\ref{kkk})
 that the factorization of dipole scattering
amplitudes is explicitly violated by a position--dependent
multiplicative factor \begin{align} \langle T_1T_2 \rangle \approx
c_{12}\langle T_1\rangle \langle T_2 \rangle \qquad c_{12}\gg 1,
\label{100}
\end{align} where $T$ is the single dipole scattering amplitude
and $\langle...\rangle$ denotes the averaging over the target
wavefunction. Eq.~(\ref{100}) should be contrasted with the fact
that
 scattering amplitudes computed in the BK--JIMWLK framework
\cite{B}  essentially factorize
\begin{align} \langle T_1T_2\rangle \approx \langle T_1 \rangle \langle
T_2 \rangle + {\cal O}\left(\frac{1}{N_c^2}\right). \label{1000}
\end{align}
 The gluon splitting diagrams which lead to Eq.~(\ref{100}) are not included
 in the BK--JIMWLK equation which rather sums gluon recombination
 diagrams. Thus it is not surprising that the 
 factorization in Eq.~(\ref{1000}) does not hold for a more
 general evolution.
  While Eq.~(\ref{1000}) may be valid if one starts with a
  large nucleus with totally
  uncorrelated partons \cite{larry} and follows the BK--JIMWLK
  evolution up to not so high energies, it is likely that the
   correlation in the transverse plane developed in the dilute regime significantly
   affects
   the nonlinear evolution of  hadrons  as in the
   case of the gluon number fluctuation \cite{IM,imm,shoshi,edmond,golec,diff}.

\section{Single dipole density}

\setcounter{equation}{0} In this section, we review the properties
of the single dipole distribution. The techniques used here
 can be directly applied to the analysis of the dipole pair density in the next section.
The single dipole density evolved up to  rapidity $Y$ is given by
\begin{align} n_Y(x_{01},x) = 2\frac{x_{01}^2}{x^2}\int
\frac{d\gamma}{2\pi i}
e^{\chi(\gamma)Y-\gamma\ln\frac{x_{01}^2}{x^2}},
\label{single}\end{align}
 $x_{01}=x_0-x_1$ and $x\equiv x_{23}=x_2-x_3$ denote
the coordinates of the parent dipole and the child dipole,
respectively.  We shall use the letter
 $x$ for two--dimensional real vectors and $z$ for corresponding complex
coordinates. By slight abuse of notation, we use $x$ also for the
magnitude of two dimensional vectors.  $\chi$ is the usual BFKL
 eigenvalue \begin{align} \chi(\gamma)=2\bar{\alpha}_s
 \mbox{Re} \{\psi(1)-\psi(\gamma)\}. \end{align} The saddle point of the
$\gamma$--integral is given by
\begin{align} \chi'(\gamma)Y=\ln \frac{x_{01}^2}{x^2} \end{align}
 When $x_{01}> x$, the saddle point is in the region
 $1>\gamma>\frac{1}{2}$, and
\begin{align} n_Y(x_{01},x) \sim
 e^{\chi(\gamma)Y}\left(\frac{x_{01}^2}{x^2}\right)^{1-\gamma}.\label{bindep}
 \end{align}
 $n$ is proportional to the  scattering amplitude between dipoles
 of sizes $x_{01}$ and $x$.
\begin{align} \label{jim} T_Y(x_{01},x)=\frac{\pi  \alpha_s^2 x^2}
{2\gamma^2 (1-\gamma)^2} n_Y(x_{01},x) \end{align}
 Eq.~(\ref{single}) is integrated over the impact parameter
 $b\equiv \frac{x_2+x_3}{2}-\frac{x_0+x_1}{2}$ between the parent and  child dipoles .
  The $b$--dependent distribution is
 \begin{align} &n_Y(x_{01},x,b)=\frac{16}{x^2} \sum_n \int
 \frac{d\nu}{(2\pi)^3}
 (\nu^2+\frac{n^2}{4})\,  e^{\chi(n,\nu)Y}
  \nonumber \\ & \quad \times \int d^2\omega
E^{1-h,1-\bar{h}}(b+\frac{x}{2}-\omega,b-\frac{x}{2}-\omega)
E^{h,\bar{h}}(\frac{x_{01}}{2}-\omega, -\frac{x_{01}}{2}-\omega).
 \label{bdep}
\end{align} $E$ is the eigenfunction of the SL(2,C) group  \begin{align}
&E^{h,\bar{h}}(x_{0\gamma},x_{1\gamma})=(-1)^n
\left(\frac{z_{01}}{z_{0\gamma}z_{1\gamma}}\right)^h
\left(\frac{\bar{z}_{01}}{\bar{z}_{0\gamma}\bar{z}_{1\gamma}}\right)^{\bar{h}},
\nonumber \\
&E^{h,\bar{h}*}(x_{0\gamma},x_{1\gamma})=
E^{1-h,1-\bar{h}}(x_{0\gamma},x_{1\gamma})
\end{align} with
  $h=\frac{1-n}{2}+i\nu$, $\bar{h}=\frac{1+n}{2}+i\nu=1-h^*$.
 Eq.~(\ref{single})
 is obtained from  Eq.~(\ref{bdep}) by integrating over $b$
\begin{align}
n_Y(x_{01},x) =\int d^2b \ n_Y(x_{01},x,b), \label{bint}
\end{align}  keeping only the $n=0$ term and identifying
 $h=\frac{1}{2}+i\nu \equiv \gamma$. The $w$--integral in
Eq.~(\ref{bdep}) has been carried out in \cite{lip,robi2}. Due to
 global conformal symmetry, the result depends only on the
anharmonic ratio
 \begin{align} \rho\equiv
 \frac{z_{01}z}{z_{02}z_{13}}, \end{align} The $n=0$ term gives,
 \begin{align}
 n_{0\nu}(x_{01},x,b)\equiv \frac{2\nu^2}{x^2\pi^4}\Bigl(b_{0,\nu}|\rho|^{2(1-\gamma)}
 F(1-\gamma,1-\gamma,
 2(1-\gamma);\rho)F(1-\gamma,1-\gamma,2(1-\gamma);\bar{\rho})\nonumber \\
 +b^*_{0,\nu}|\rho|^{2\gamma}
 F(\gamma,\gamma,
 2\gamma;\rho)F(\gamma,\gamma,2\gamma;\bar{\rho})\Bigr), \label{lev} \end{align}
 where $F$ is the hypergeometric function and \begin{align}
 b_{0,\nu}=\pi^3
\frac{2^{4i\nu}}{i\nu}\frac{\Gamma(\frac{1}{2}-i\nu)\Gamma(1+i\nu)}
 {\Gamma(\frac{1}{2}+i\nu)\Gamma(1-i\nu)}.
 \end{align}
  Consider the case $x_{01}\gg x$ and look at the region
  of small impact parameters $x_{01}\gg b$. In this region,
 \begin{align} \rho \approx \frac{-4z}{z_{01}}, \qquad |\rho|\ll 1,\end{align} and one may approximate
  $F(...;\rho) \approx 1$. We obtain\footnote{Ref.~\cite{wallon} uses the
  following approximation
  \begin{align} b_{0,\nu}
  \approx \pi^3 \frac{16^{2i\nu}}{i\nu}. \end{align} This is valid as long as $\nu$ is close to zero and
  leads to a factor $\left(\frac{ 16 x^2_{01}}{x^2}\right)^{1-\gamma}$.
  However, in our case the
  saddle point for $\nu$ is not assumed to be small, but rather determined from
   external parameters (dipole sizes).}
 \begin{align} n_Y(x_{01},x,b)=\int
\frac{d\nu}{2\pi}n_{0,\nu}(x_{01},x,b)e^{\chi(0,\nu)Y}\approx
\frac{1}{ x_{01}^2}\int d\nu e^{\chi(0,\nu) Y}
\frac{16^{\gamma}\nu^2b_{0,\nu}^{*}}{\pi^5}\left(\frac{
x^2_{01}}{x^2}\right)^{1-\gamma}+c.c. \label{double}
\end{align}
 Comparing Eq.~(\ref{single}) and Eq.~(\ref{double}), one sees
 that in the saddle point approximation,
  \begin{align} n_Y(x_{01},x)\sim x_{01}^2
 \, n_Y(x_{01},x,b\ll x_{01}). \label{point} \end{align}
Therefore, roughly child
 dipoles are uniformly distributed inside the area
 $x_{01}^2$ (c.f., Eq.~(\ref{bint})).
 On the other hand, the dipole density at large impact parameters $b \gg x_{01}$ are
 suppressed. Indeed, in this region,
  $\rho \approx \frac{z_{01}z}{b^2}$, and
  \begin{align}  n_Y(x_{01},x,b) \approx
   \frac{1}{x^2}\int d\nu
  \frac{\nu^2b_{0,\nu}^*}{\pi^5}\left(\frac{x_{01}^2x^2}{b^4}
  \right)^{\gamma} e^{\chi(0,\nu) Y}+c.c.. \label{b}
   \end{align} At the saddle point, $n(b)\sim 1/b^{4\gamma}$ where
   $\gamma$ is determined from
   $\chi'(\gamma)Y=\ln\frac{b^4}{x_{01}^2x^2}$.

 Let us compare this $b$--dependence with that of the saturation momentum.
  The dipole--dipole scattering amplitude at a fixed
  impact parameter $b$ is \begin{align} T_Y(x_{01},x,b)=\int
  d^2b'\frac{d^2x'}{2\pi
  x'^2}A_0(x,x',b-b')n_Y(x_{01},x',b'), \end{align} where $A_0$ is the
  dipole--dipole scattering amplitude in the two--gluon exchange approximation. Since
  $A_0(b-b')$ decays like $1/(b-b')^4$,  one may approximate
  \begin{align}T_Y(x_{01},x,b)\approx \int
  d^2b'A_0(x,x',b-b') \int \frac{d^2x'}{2\pi
  x'^2}n_Y(x_{01},x',b)\nonumber \\
  =\pi \alpha_s^2\int \frac{d^2x'}{2\pi
  x'^2}r^2_<(1+\ln \frac{r_>}{r_<})n_Y(x_{01},x',b) \label{approximation} \end{align}
  where $r_<=\mbox{min}\{x,x'\}$ and $r_>=\mbox{max}\{x,x'\}$. Using the large $b$ form of $n$, Eq.~(\ref{b}), one
  obtains \footnote{ Compare with Eq.~(\ref{jim}). The factor 2 difference in the
denominator
  is due to the definition \begin{align}
   \frac{T_Y(t=0)}{2}=\int d^2bT_Y(b).\end{align}}
   \begin{align}
   &T_Y(x_{01},x,b)\approx \frac{\pi \alpha_s^2 x^2}{4\gamma^2(1-\gamma)^2}
  n_Y(x_{01},x,b). \end{align}

  The local ($b$--dependent) saturation momentum can be determined
  by the constancy of the
  exponential factor of $T$ in the integral representation along
  the line $x=1/Q_s(b,Y)$ \cite{geom,dio},
  and reads \begin{align} Q^2_s(b,Y)\sim \frac{x^2_{01}}{ b^4}e^{\frac{\chi(\gamma_s)}
   {\gamma_s}Y}, \label{iij}\end{align} where $\gamma_s\approx 0.628$ solves  $\chi'(\gamma_s)=\frac{\chi(\gamma_s)}{\gamma_s}$.
  Repeating the same procedure for $x_{01}\gg b$, we get \begin{align}
  Q^2_s(b,Y)\sim \frac{1}{ x_{01}^2}e^{\frac{\chi(\gamma_s)}
   {\gamma_s}Y}. \label{qsat} \end{align}

   If we take $x$ to be close to the saturation
   line $\sim 1/Q_s(b,Y)$,
   $\gamma\approx \gamma_s$, and  the geometric scaling
   \cite{geom,dio,traveling}
   holds locally in the two
    ($b\gg x_{01}$, $b\ll x_{01}$) regimes \cite{mun}
    \begin{align} T_Y(x_{01},x,b)\sim x^2 n_Y(x_{01},x,b) \sim
  (x^2Q_s^2(b,Y))^{\gamma_s}.
  \label{liou} \end{align}

\section{Dipole pair density}
\setcounter{equation}{0}

\begin{figure}
\begin{center}
\centerline{\epsfig{file=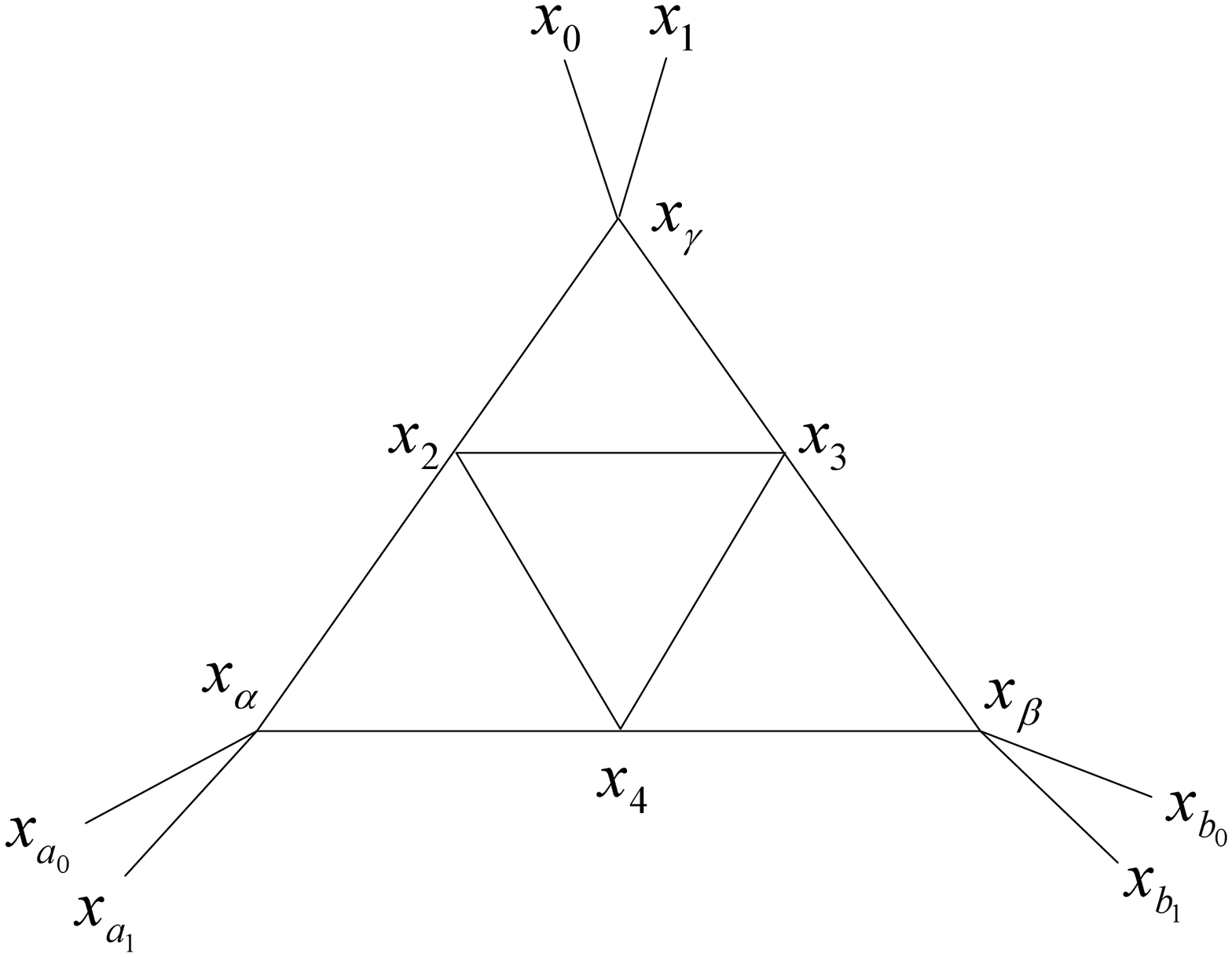,height=8.cm,width=10cm}}
\caption{\sl Graphical representation of Eq.~(\ref{im}).
 \label{fig2}}
\end{center}
\end{figure}

\begin{figure}
\begin{center}
\centerline{\epsfig{file=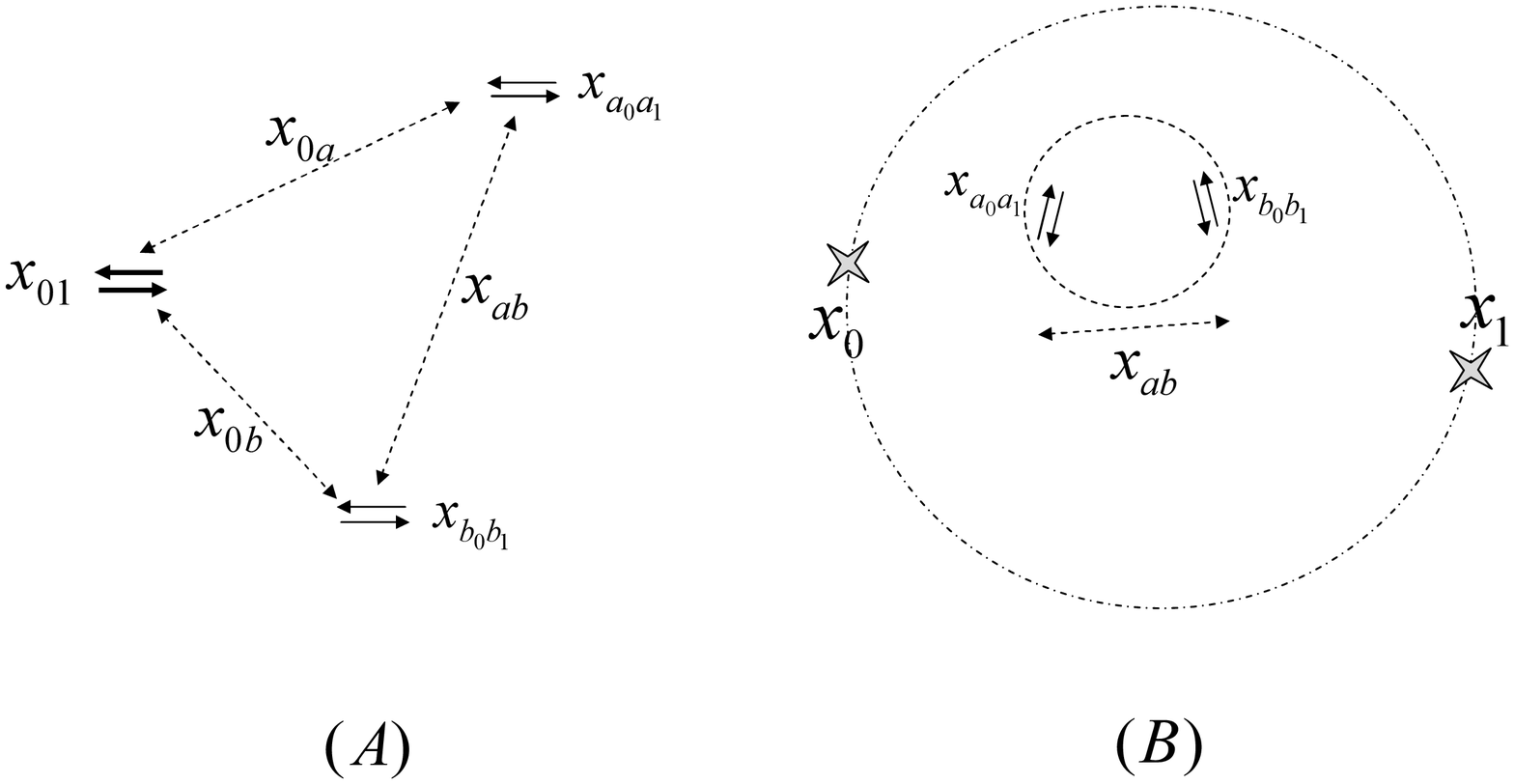,height=7.cm,width=13cm}}
\caption{\sl Case (A): All dipoles are small compared with their
mutual separations. Case (B): Child dipoles are deep inside the
large parent dipole.
 \label{fig3}}
\end{center}
\end{figure}

In this and the next section, we analyze the dipole pair density
$n^{(2)}$ in two different ways. We start with the exact
expression for the pair density as derived in \cite{1} (see, also,
\cite{vacca}).
\begin{align}
&n^{(2)}_Y(x_{01},x_{a_0a_1},x_{b_0b_1})=\int dh dh_a dh_b
\frac{1}{2x^2_{a_0a_1}x_{b_0b_1}^2}\int_0^{Y} dy\
e^{\chi(h)y+(\chi(h_a)+\chi(h_b))(Y-y)}
 \nonumber \\ &\times \int d^2x_\alpha d^2x_\beta
d^2x_\gamma
E^{h,\bar{h}}(x_{0\gamma},x_{1\gamma})E^{h_a,\bar{h}_a}
 (x_{a_0\alpha},x_{a_1\alpha})
E^{h_b,\bar{h}_b}(x_{b_0\beta},x_{b_1\beta}) \nonumber \\
&\times \int \frac{d^2x_2 d^2x_3 d^2
x_4}{x_{23}^2x_{34}^2x_{42}^2} E^{h,\bar{h}*}(x_{2 \gamma},x_{3
\gamma})E^{h_a,\bar{h}_a*} (x_{2\alpha},x_{4\alpha})
E^{h_b,\bar{h}_b*}(x_{3\beta},x_{4\beta}), \label{im}
 \end{align}
 where
 $(x_{a_0},x_{a_1})$ and $(x_{b_0},x_{b_1})$ are  coordinates of
 the child dipoles of interest. We introduced a compact notation \begin{align} \int dh \equiv \sum_n \int d\nu
 \frac{2\nu^2+n^2/2}{\pi^4}. \end{align}
A graphical representation of the coordinate integrals is shown in
Fig.~\ref{fig2}.
We will be interested in configurations where the two child
dipoles are small (typically of the order of the inverse
saturation scale $\sim 1/Q_s$) and far away from each other,
\begin{align} x_{ab}=x_a-x_b\equiv
\frac{x_{a_0}+x_{a_1}}{2}-\frac{x_{b_0}+x_{b_1}}{2} \gg
x_{a_0a_1}, \, x_{b_0b_1}, \end{align}
 and try to extract the leading $x_{ab}$ dependence of $n^{(2)}$.
 This leaves us with two interesting (and in fact, tractable) situations (see, Fig.~\ref{fig3}):
 (A) The parent
 dipole is also small $x_{01}\sim x_{a_0a_1},x_{b_0b_1} \ll x_{ab}$. (B) The
 parent dipole is large $x_{01}\gg x_{ab}\gg x_{a_0a_1},x_{b_0b_1}$.

 Let us first consider the case A. In the next section we will discuss both cases
  in a unified way.
The last line in Eq.~(\ref{im}) is the triple pomeron vertex in
perturbative QCD \cite{bar}  at large $N_c$. It has the form
\begin{align}
f(\bar{h},\bar{h}_a,\bar{h}_b)\frac{1}{(z_{\alpha}-z_\beta)^{1+h-h_a-h_b}
(z_\beta-z_\gamma)^{1+h_a-h_b-h}(z_\gamma
-z_\alpha)^{1+h_b-h-h_a}}\nonumber
\\ \times \frac{1}{(\bar{z}_{\alpha}-\bar{z}_\beta)^{1+
\bar{h}-\bar{h}_a-\bar{h}_b}
(\bar{z}_\beta-\bar{z}_\gamma)^{1+\bar{h}_a-\bar{h}_b-\bar{h}}(\bar{z}_\gamma
-\bar{z}_\alpha)^{1+\bar{h}_b-\bar{h}-\bar{h}_a}}. \label{wi}
\end{align} This structure follows immediately by noting that the
last line of Eq.~(\ref{im}) and Eq.~(\ref{wi}) transform in the
same way under the SL(2,C) transformations of $z_\alpha, \
z_\beta, \ z_\gamma$
\begin{align} z\to \frac{\alpha z+\beta}{\gamma z+\delta} \qquad (\alpha\delta-\beta\gamma=1).
\end{align}
The function $f(h,h_a,h_b)$ can be found in \cite{bia,kor}. Next
we turn to the remaining integrals $d^2x_\alpha d^2x_\beta
d^2x_\gamma$ in (\ref{im}). Since all dipoles (parent, children)
are assumed to be very small, typically $x_{0a},\, x_{0b},\,
x_{ab} \gg x_{01},\, x_{a_0a_1},\, x_{b_0b_1}$ and we may make
approximations
\begin{align} \left(\frac{z_{01}}{z_{0\gamma}z_{1\gamma}}\right)^h
\approx \left(\frac{z_{01}}{z^2_{0\gamma}}\right)^h, \ \
\left(\frac{z_{a_0a_1}}{z_{a_0\alpha}z_{a_1\alpha}}\right)^h
\approx \left(\frac{z_{a_0a_1}}{z^2_{a\alpha}}\right)^h, \ \
\left(\frac{z_{b_0b_1}}{z_{b_0\beta}z_{b_1\beta}}\right)^h \approx
\left(\frac{z_{b_0b_1}}{z^2_{b\beta}}\right)^h.
\label{appro}\end{align}  We will see later that with this
replacement one makes a mistake in the overall factor of $n^{(2)}$
by 8. After this approximation, we are left with the integral
\begin{align} \int d^2x_\alpha d^2x_\beta
d^2x_\gamma
\frac{1}{z^{2h}_{0\gamma}z^{2h_a}_{a\alpha}z^{2h_b}_{b\beta}}\frac{1}
{z_{\alpha\beta}^{1+h-h_a-h_b}
z_{\beta\gamma}^{1+h_a-h_b-h}z_{\gamma\alpha}
^{1+h_b-h-h_a}}\nonumber
\\ \times \frac{1}{\bar{z}^{2\bar{h}}_{0\gamma}\bar{z}^{2\bar{h}_a}_{a\alpha}
\bar{z}^{2\bar{h}_b}_{b\beta}}\frac{1}{\bar{z}_{\alpha\beta}^{
1+\bar{h}-\bar{h}_a-\bar{h}_b}
\bar{z}_{\beta\gamma}^{1+\bar{h}_a-\bar{h}_b-\bar{h}}\bar{z}_{\gamma\alpha}
^{\bar{h}_b-\bar{h}-\bar{h}_a}}. \label{long}
\end{align} One can check that this integral transforms in the
same way under the SL(2,C) transformation of $z_0,\ z_a,\ z_b$ as
\begin{align}
\frac{1}{z_{0a}^{h+h_a-h_b}z_{0b}^{h+h_b-h_a}z_{ab}^{h_a+h_b-h}}
\frac{1}{\bar{z}_{0a}^{\bar{h}+\bar{h}_a-\bar{h}_b}\bar{z}_{0b}^{\bar{h}+
\bar{h}_b-\bar{h}_a}\bar{z}_{ab}^{\bar{h}_a+\bar{h}_b-\bar{h}}}.
\label{imp}
\end{align}

The coefficient can be easily obtained. In the dominant case of
$n=n_a=n_b=0$ where $h=\bar{h}=\frac{1}{2}+i\nu \equiv \gamma$,
$h_{a}=\bar{h}_a\equiv \gamma_a$, $h_b=\bar{h}_b\equiv \gamma_b$,
(Generalization to the case $h\neq \bar{h}$ is straightforward.)

 \begin{align}
\int d^2x_\alpha d^2x_\beta d^2x_\gamma
\frac{1}{|z_{0\gamma}|^{4\gamma}|z_{a\alpha}|^{4\gamma_a}|z_{b\beta}|^{4\gamma_b}}
\frac{1} {|z_{\alpha\beta}|^{2(\gamma-\gamma_a-\gamma_b+1)}
|z_{\beta\gamma}|^{2(\gamma_a-\gamma_b-\gamma+1)}|z_{\gamma\alpha}|^{2(\gamma_b-
\gamma-\gamma_a+1)}} \nonumber \\
=\frac{g(\gamma,\gamma_a,\gamma_b)}{|z_{0a}|^{2(\gamma+\gamma_a-\gamma_b)}
|z_{0b}|^{2(\gamma+\gamma_b-\gamma_a)}|z_{ab}|^{2(\gamma_a+\gamma_b-\gamma)}}
\end{align} When $\gamma>1/2$,  a pole at $z_\gamma=z_0$ is not integrable. The
following result should be regarded as analytic continuation from
convergent values of $\gamma$'s. Using a conformal transformation,
one can set $z_a=0$, $z_b=1$, $z_0=\infty$.
\begin{align}
 g(\gamma,\gamma_a,\gamma_b)= \int d^2x_\alpha d^2x_\beta d^2x_\gamma
\frac{1}{|z_{\alpha}|^{4\gamma_a}|1-z_{\beta}|^{4\gamma_b}}\frac{1}
{|z_{\alpha\beta}|^{2(\gamma-\gamma_a-\gamma_b+1)}
|z_{\beta\gamma}|^{2(\gamma_a-\gamma_b-\gamma+1)}
|z_{\gamma\alpha}|^{2(\gamma_b-\gamma-\gamma_a+1)}}
\end{align} Evaluating the integrals in the order of $d^2x_\gamma$,
$d^2x_\beta$ and $d^2x_\alpha$,
 one obtains
\begin{align} &g(\gamma,\gamma_a,\gamma_b)
=\pi^3\frac{\Gamma(1-2\gamma)\Gamma(1-2\gamma_a)\Gamma(1-2\gamma_b)\Gamma(\gamma+\gamma_a+\gamma_b-1)
 }
{\Gamma(2\gamma)\Gamma(2\gamma_a)\Gamma(2\gamma_b)\Gamma(2-\gamma-\gamma_a-\gamma_b)
}\nonumber \\
&\times
\frac{\Gamma(\gamma+\gamma_a-\gamma_b)\Gamma(\gamma+\gamma_b-\gamma_a)
\Gamma(\gamma_a+\gamma_b-\gamma)}
{\Gamma(1+\gamma_b-\gamma_a-\gamma)\Gamma(1+\gamma_a-\gamma_b-\gamma)
\Gamma(1+\gamma-\gamma_a-\gamma_b)},
\end{align}
 and therefore, \begin{align} & n^{(2)} =\int d\gamma d\gamma_a d\gamma_b
\frac{1}{2x_{a_0a_1}^2x_{b_0b_1}^2}g(\gamma,\gamma_a,\gamma_b)f(\bar{\gamma},\bar{\gamma}_a,\bar{\gamma}_b)\int_0^{Y}
dy \frac{x_{01}^{2\gamma}
 x_{a_0a_1}^{2\gamma_a}x_{b_0b_1}^{2\gamma_b}\, e^{\chi(\gamma)y+(\chi(\gamma_a)+\chi(\gamma_b))(Y-y)}}{x_{0a}^{2(\gamma+\gamma_a-\gamma_b)}x_{0b}^{2(\gamma+\gamma_b-\gamma_a)}
x_{ab}^{2(\gamma_a+\gamma_b-\gamma)}} \nonumber \\ & =\int d\gamma
d\gamma_a d\gamma_b
\frac{1}{2x_{a_0a_1}^2x_{b_0b_1}^2}g(\gamma,\gamma_a,\gamma_b)
f(\bar{\gamma},\bar{\gamma_a},\bar{\gamma_b})\int_0^{Y}
dy \, \nonumber \\
& \times
\exp\left(\chi(\gamma)y+(\chi(\gamma_a)+\chi(\gamma_b))(Y-y)-\gamma\ln
\left(\frac{x_{0a}^2 x_{0b}^2}{x_{01}^2x_{ab}^2}\right) - \gamma_a
\ln \left( \frac{x_{0a}^2x_{ab}^2}{x^2_{a_0a_1}x_{0b}^2}\right) -
\gamma_b \ln
\left(\frac{x_{0b}^2x_{ab}^2}{x^2_{b_0b_1}x_{0a}^2}\right) \right). \label{!} \nonumber \\
 \end{align}

 The remaining integrals  may be evaluated in the
 saddle point approximation.
 The saddle points for $y,\ \gamma, \ \gamma_{a,b}$  are given by
 the solution to
 \begin{align}
 \chi(\gamma)=\chi(\gamma_a)+\chi(\gamma_b), \label{min}\\
 \chi'(\gamma_a)(Y-y)=\ln \frac{x_{0a}^2x_{ab}^2}{x_{0b}^2x^2_{a_0a_1}} \gg 1, \label{as}\\
 \chi'(\gamma_b)(Y-y)=\ln \frac{x_{0b}^2x_{ab}^2}{x_{0a}^2x^2_{b_0b_1}} \gg 1, \\
 \chi'(\gamma)y=\ln \frac{x_{0a}^2x_{0b}^2}{x_{ab}^2x_{10}^2} \gg 1, \label{nao}\end{align}
 and the dipole pair density behaves like
  \begin{align}
  n^{(2)}_Y(x_{01},x_{a_0a_1},x_{b_0b_1})\sim
  \frac{e^{(\chi(\gamma_a)+\chi(\gamma_b))Y}}
  {x_{a_0a_1}^2x_{b_0b_1}^2}
  \frac{x_{01}^{2\gamma}
 x_{a_0a_1}^{2\gamma_a}x_{b_0b_1}^{2\gamma_b}}{x_{0a}^{2(\gamma+\gamma_a-\gamma_b)}x_{0b}^{2(\gamma+\gamma_b-\gamma_a)}
x_{ab}^{2(\gamma_a+\gamma_b-\gamma)}}.
  \label{222}
 \end{align} The $x_{01}^{2\gamma}$ behavior of $n^{(2)}$ was pointed out
 in \cite{al}. (See, Eq.~(A.2) of \cite{al}.) From Eq.~(\ref{nao})
 we see that $\frac{1}{2} < \gamma <1$,
 and this justifies the conjecture below Eq.~(A.7) of \cite{al}.
 The factor $1/x_{0a}^{2\gamma}$ (with $\gamma_a=\gamma_b=\frac{1}{2}$) was
 found in \cite{bialas} in the context of dipole production at large
 transverse distances.

 The factor $x_{ab}^{-2(\gamma_a+\gamma_b-\gamma)}$ characterizes the correlation
 of dipoles in impact parameter space. To see the
  significance of this factor, consider
 scattering of two
 dipoles $x_{a_0a_1},x_{b_0b_1}$ on a target dipole $x_{01}$ at large impact
 parameter. The scattering amplitude is given by
 \begin{align} & T_Y(x_{01};x_{a_0a_1},x_{b_0b_1})=\int \frac{d^2x}{2\pi
 x^2}\frac{d^2x'}{2\pi x'^2} \int d^2b d^2b' n^{(2)}_Y(x_{01},
 xb,x'b') \nonumber \\ & \qquad \qquad \qquad \qquad \qquad \times
 A(x_{a_0a_1}x,x_{a}-\frac{x_0+x_1}{2}-b)A(x_{b_0b_1}x',x_{b}-\frac{x_0+x_1}{2}-b')\nonumber \\
& \qquad \qquad \qquad \qquad \approx
  \frac{ \pi^2 \alpha_s^4x_{a_0a_1}^2x_{b_0b_1}^2}{16\gamma_a^2(1-\gamma_a)^2
  \gamma_b^2(1-\gamma_b)^2} n^{(2)}_Y(x_{01},
 x_{a_0a_1},x_{b_0b_1})\nonumber \\
 & \qquad \qquad \qquad \qquad \sim
 T_Y(x_{01},x_{a_0a_1},x_{0a})T_Y(x_{01},x_{b_0b_1},x_{0b})
 \left(\frac{x_{0a}x_{0b}}{x_{01}x_{ab}}\right)^{2(\gamma_a+\gamma_b-\gamma)}.
 \label{ka} \end{align}
In the second line, we have used the same approximation as in
Eq.~(\ref{approximation}). The last line should be taken with care
since $\gamma_a$ as determined from Eq.~(\ref{as}) does not
coincide with the anomalous dimension of
$T(x_{01},x_{a_0a_1},x_{0a})$, the latter being determined from
$\chi'(\gamma_a)Y=\ln \frac{x_{0a}^4}{x_{a_0a_1}^2x_{01}^2}$
(c.f., Eqs.~(\ref{as}), (\ref{nao}) and note that $\gamma\neq
\gamma_a$). Even if we neglect this difference, we see from
Eq.~(\ref{ka}) that the factorization of two--dipole amplitude is
explicitly violated by a nontrivial position--dependent
factor.\footnote{ We note that the two--dipole scattering
amplitude that appears on the right hand side of the BK--JIMWLK
equation \cite{B} is for \emph{contiguous} dipoles,
$x_{a_1}=x_{b_0}$. Our present approach does not apply to this
interesting case since we assumed $x_{ab}\gg
x_{a_0a_1},x_{b_0b_1}$.}
\begin{align}
\left(\frac{x_{0a}x_{0b}}{x_{01}x_{ab}}\right)^{2(\gamma_a+\gamma_b-\gamma)}
\gg 1.
\end{align}

\section{Improved calculation}
\setcounter{equation}{0}

Let us return to the integral appearing in Eq.~(\ref{im}).
\begin{align} & I= \int d^2x_\alpha d^2x_\beta d^2x_\gamma
E^{h,\bar{h}}(x_{0\gamma},x_{1\gamma})E^{h_a,\bar{h}_a}
 (x_{a_0\alpha},x_{a_1\alpha})
E^{h_b,\bar{h}_b}(x_{b_0\beta},x_{b_1\beta}) \nonumber \\
&\times \int \frac{d^2x_2d^2x_3d^2x_4}{x_{23}^2x_{34}^2x_{42}^2}
E^{h,\bar{h}*}(x_{2 \gamma},x_{3 \gamma})E^{h_a,\bar{h}_a*}
(x_{2\alpha},x_{4\alpha})
E^{h_b,\bar{h}_b*}(x_{3\beta},x_{4\beta}).
 \end{align}  Instead of first integrating over $x_{2,3,4}$ (`reggeon coordinates')
  as we did before, now we
 integrate over $x_{\alpha,\beta,\gamma}$
 (`Pomeron coordinates', see, Fig.~\ref{fig2}) first.
  \begin{align}
 I=\frac{1}{(2\pi^2)^3} \int \frac{d^2x_2d^2x_3d^2x_4}{x_{23}^2x_{34}^2x_{42}^2}
  \Bigl(b^*_{0,\nu}|\rho|^{2\gamma}
 F(\gamma,\gamma,
 2\gamma;\rho)F(\gamma,\gamma,2\gamma;\bar{\rho})+c.c.\Bigr)
 \nonumber \\ \times
 \Bigl(b^*_{0,\nu_a}|\rho_a|^{2\gamma_a}
 F(\gamma_a,\gamma_a,
 2\gamma_a;\rho_a)F(\gamma_a,\gamma_a,2\gamma_a;\bar{\rho}_a)+c.c.\Bigr)
 \nonumber \\\times \Bigl(b^*_{0,\nu_b}|\rho_b|^{2\gamma_b}
 F(\gamma_b,\gamma_b,
 2\gamma_b;\rho_b)F(\gamma_b,\gamma_b,2\gamma_b;\bar{\rho}_b)+c.c.\Bigr),
 \label{label}
 \end{align}
 where \begin{align} \rho\equiv
 \frac{z_{01}z_{23}}{z_{02}z_{13}},\ \ \rho_a\equiv
 \frac{z_{a_0a_1}z_{24}}{z_{a_02}z_{a_14}}, \ \ \rho_b\equiv
 \frac{z_{b_0b_1}z_{34}}{z_{b_03}z_{b_14}},\end{align} are anharmonic ratios. By assumption,
 $\rho_a$ and $\rho_b$ are small, and one may approximate
 $F(...;\rho_{a,b})\approx 1$. A remarkable point is that $\rho$
 is small \emph{both} in the limits of $x_{01}\to\infty$ and $x_{01}\to
 0$ and one may approximate $F(...;\rho)\approx 1.$ Expanding the brackets,
 we get eight terms. The first
 term reads \begin{align}  I_1=\frac{1}{8\pi^6} b^*_{0,\nu}b^*_{0,\nu_a}b^*_{0,\nu_b}
 \int \frac{d^2x_2d^2x_3d^2x_4}{x_{23}^2x_{34}^2x_{42}^2} \left(\frac{x_{01}x_{23}}{x_{02}x_{13}}\right)^{2\gamma}
 \left(\frac{x_{a_0a_1}x_{24}}{x_{a2}x_{a4}}\right)^{2\gamma_a}
\left(\frac{x_{b_0b_1}x_{34}}{x_{b3}x_{b4}}\right)^{2\gamma_b}
\label{ui}
  \end{align}
   If we take the limit $x_{01}\to 0$, this is the same
  integral which gives the triple Pomeron vertex $f(\gamma,\gamma_a,\gamma_b)$.
 \begin{align} &\int \frac{d^2x_2d^2x_3d^2x_4}{x_{23}^2x_{34}^2x_{42}^2} \left(\frac{x_{01}x_{23}}{x_{02}x_{13}}\right)^{2\gamma}
 \left(\frac{x_{a_0a_1}x_{24}}{x_{a2}x_{a4}}\right)^{2\gamma_a}
\left(\frac{x_{b_0b_1}x_{34}}{x_{b3}x_{b4}}\right)^{2\gamma_b}
\nonumber \\ &\approx x_{01}^{2\gamma}x_{a_0a_1}^{2\gamma_a}
x_{b_0b_1}^{2\gamma_b}\int
\frac{d^2x_2d^2x_3d^2x_4}{x_{23}^2x_{34}^2x_{42}^2}
\left(\frac{x_{23}}{x_{02}x_{03}}\right)^{2\gamma}
 \left(\frac{x_{24}}{x_{a2}x_{a4}}\right)^{2\gamma_a}
\left(\frac{x_{34}}{x_{b3}x_{b4}}\right)^{2\gamma_b} \nonumber \\
&=x_{01}^{2\gamma}x_{a_0a_1}^{2\gamma_a}
x_{b_0b_1}^{2\gamma_b}\frac{f(\gamma,\gamma_a,\gamma_b)}{x_{0a}^{2(\gamma+\gamma_a-\gamma_b)}
x_{0b}^{2(\gamma+\gamma_b-\gamma_a)}x_{ab}^{2(\gamma_a+\gamma_b-\gamma)}}.
\label{limit}
  \end{align}
 For a later use, we note that when  $\gamma_a=\gamma_b$,
 \begin{align}
 I_1=\frac{1}{8\pi^6} b^*_{0,\nu}b^{*2}_{0,\nu_a}
 f(\gamma,\gamma_a,\gamma_a)\left(\frac{x_{a_0a_1}x_{b_0b_1}}{x_{ab}^2}\right)^{2\gamma_a}
 \left(\frac{x_{01}x_{ab}}{x_{0a}x_{0b}}\right)^{2\gamma}, \quad
 (x_{01}\to 0).
 \label{later}
 \end{align}
  Eq.~(\ref{limit}) should coincide with our previous
  result Eq.~(\ref{!}), so we obtain an identity  \begin{align}
   g(\gamma,\gamma_a,\gamma_b)f(\bar{\gamma},\bar{\gamma}_a,\bar{\gamma}_b)
   =\frac{1}{8\pi^6}b^*_{0,\nu}b^*_{0,\nu_a}b^*_{0,\nu_b}f(\gamma,\gamma_a,\gamma_b).
   \label{kor} \end{align} Eq.~(\ref{kor}) is a straightforward  generalization of
   the relation between $f(\gamma,\gamma_a,\gamma_b)$ and $f(\bar{\gamma},\gamma_a,\gamma_b)$
   derived in \cite{kor}.
   We also see that the
previous approximation Eq.~(\ref{appro}) misses
    the seven other terms in Eq.~(\ref{label}) which contribute
    equally to $n^{(2)}$ due to the symmetry $\gamma \leftrightarrow 1-\gamma$ of
   the integrals $\int d\gamma
    d\gamma_a d\gamma_b$. We take this into account by multiplying Eq.~(\ref{ui})
    by 8. \begin{align}  I\to \frac{1}{\pi^6} b^*_{0,\nu}b^*_{0,\nu_a}b^*_{0,\nu_b}
 \int \frac{d^2x_2d^2x_3d^2x_4}{x_{23}^2x_{34}^2x_{42}^2}\left(\frac{x_{01}x_{23}}{x_{02}x_{13}}\right)^{2\gamma}
 \left(\frac{x_{a_0a_1}x_{24}}{x_{a2}x_{a4}}\right)^{2\gamma_a}
\left(\frac{x_{b_0b_1}x_{34}}{x_{b3}x_{b4}}\right)^{2\gamma_b}
\label{li}
  \end{align}
 Now we would like to evaluate this for $x_{01}\to \infty$. In the following,
 we assume that  $\gamma_a=\gamma_b$, which will be approximately
 valid at the saddle point when $x_{a_0a_1}\sim x_{b_0b_1}$.
  Then
Eq.~(\ref{li}) takes the form \begin{align} I=\frac{1}{\pi^6}
b^*_{0,\nu}b^{*2}_{0,\nu_a}\left(\frac{x_{a_0a_1}x_{b_0b_1}}{x^2_{ab}}\right)^{2\gamma_a}
\int \frac{d^2x_2d^2x_3d^2x_4}{x_{23}^2x_{34}^2x_{42}^2}
\left(\frac{x_{01}x_{23}}{x_{02}x_{13}}\right)^{2\gamma}
\left(\frac{x_{24}x_{34}x_{ab}^2}{x_{a2}x_{a4}x_{b3}x_{b4}}\right)^{2\gamma_a}.
\end{align}
It is easy to see  from the SL(2,C) invariance that the result of
the integration must have the structure [In fact, this property
holds only for $\gamma_a=\gamma_b$.] \begin{align}
I=\frac{1}{\pi^6}
b^*_{0,\nu}b^{*2}_{0,\nu_a}\left(\frac{x_{a_0a_1}x_{b_0b_1}}{x^2_{ab}}\right)^{2\gamma_a}
h\left(\rho', \bar{\rho}'\right), \end{align} where
\begin{align} \rho'= \frac{z_{01}z_{ab}}{z_{0a}z_{1b}},
\end{align} is the anharmonic ratio of the external points.
 In the limit $x_{01}\to 0$, $\rho' \to 0$ and $h$
 should reproduce Eq.~(\ref{later}). \begin{align}
 h(\rho',\bar{\rho}')\approx
 f(\gamma,\gamma_a,\gamma_a)|\rho'|^{2\gamma}, \qquad (\rho' \to 0). \end{align}
 Our observation is that the limit $x_{01} \to \infty$ also leads
 to $\rho'\to 0$.  Therefore, when $x_{01} \to \infty$,
\begin{align} I\approx  \frac{1}{\pi^6}
b^{*}_{0,\nu}b^{*2}_{0,\nu_a}
f(\gamma,\gamma_a,\gamma_a)\left(\frac{x_{a_0a_1}x_{b_0b_1}}{x^2_{ab}}\right)^{2\gamma_a}\left(
\frac{x_{01}x_{ab}}{x_{0a}x_{1b}}\right)^{2\gamma}\nonumber \\
=8g(\gamma,\gamma_a,\gamma_a)f(\bar{\gamma},\bar{\gamma}_a,\bar{\gamma}_a)
\left(\frac{x_{a_0a_1}x_{b_0b_1}}{x^2_{ab}}\right)^{2\gamma_a}\left(
\frac{x_{01}x_{ab}}{x_{0a}x_{1b}}\right)^{2\gamma},
\end{align} and we obtain the behavior of $n^{(2)}$ at the saddle
point
 \begin{align}
n^{(2)}_Y(x_{0}x_1;x_{a_0}x_{a_1},x_{b_0}x_{b_1})\sim
 \frac{1}{x^2_{a_0a_1}x_{b_0b_1}^2}
e^{2\chi(\gamma_a)Y}b^{*}_{0,\nu}b^{*2}_{0,\nu_a}
f(\gamma,\gamma_a,\gamma_a)
\left(\frac{x_{a_0a_1}x_{b_0b_1}}{x^2_{ab}}\right)^{2\gamma_a}\left(
\frac{x_{01}x_{ab}}{x_{0a}x_{1b}}\right)^{2\gamma}, \label{val}
 \end{align}
 with $\gamma$ and $\gamma_a$ (and also $y$) determined
 from the saddle point equations
 \begin{align}
& \chi(\gamma)=2\chi(\gamma_a), \label{15} \\
& \chi'(\gamma)y=\ln \frac{x_{0a}^2x_{1b}^2}{x_{01}^2x_{ab}^2} \gg 1,\\
& \chi'(\gamma_a)(Y-y)=\ln \frac{x_{ab}^2}{x_{a_0a_1}^2} \gg 1.
\label{17} \end{align} Note that $\frac{1}{2} < \gamma_a < \gamma
< 1$.

  Eq.~(\ref{val}) is valid \emph{both }for $x_{01} \ll x_{ab}$ and
 $x_{01} \gg x_{ab}$. Due to conformal symmetry, the two cases of Fig.~\ref{fig3}
  are mathematically identical. In the latter case, if the dipole $x_{ab}$ is deeply inside
 the parent dipole $x_{01}$, one can make an approximation
\begin{align}
\left(\frac{x_{01}x_{ab}}{x_{0a}x_{1b}}\right)^{2\gamma}\approx
\left( \frac{x_{ab}}{x_{01}}\right)^{2\gamma}.
\end{align}
 It is interesting to note that in the same limit we may rewrite $n^{(2)}$ as
\begin{align} n^{(2)}_Y \approx \frac{f(\gamma,\gamma_a,\gamma_a)}
{x_{a_0a_1}^2x_{b_0b_1}^2}\int^Y_0 dy \frac{x_{ab}^2}{x_{01}^2}
n_y(x_{01},x_{ab}) \frac{ x_{a_0a_1}^2}{x_{ab}^2}
n_{Y-y}(x_{ab},x_{a_0a_1}) \frac{x_{b_0b_1}^2}{x_{ab}^2}
n_{Y-y}(x_{ab},x_{b_0b_1}). \label{www}
\end{align}
 Eq.~(\ref{www}) provides an intuitive understanding of the
 result. The parent dipole $x_{01}$ emits a child dipole $x_{ab}$  inside
 the area $x_{01}^2$ with uniform probability (c.f., Eq.~(\ref{point})).
 The geometrical factor $x_{ab}^2/x_{01}^2 $ specifies the
 location of the dipole $x_{ab}$.
  Then the dipole $x_{ab}$ splits into two
 dipoles of similar size $\sim x_{ab}$ through the triple Pomeron vertex,
 $f(\gamma,\gamma_a,\gamma_a)$. Finally, each of the two dipoles  emits
 a child dipole of size $x_{a_0a_1}$ (or $x_{b_0b_1}$)
  inside the area $\sim x_{ab}^2$ again with uniform probability, and the
 two child dipoles $x_{a_0a_1}$ and $x_{b_0b_1}$ roughly fall within a distance $x_{ab}$
 (see, Fig.~\ref{fig3}B). Another representation of
 $n^{(2)}$ is (c.f., Eq.~(\ref{jim})) \begin{align} n^{(2)}_Y \propto \int^Y_0 dy
T_y(x_{01},x_{ab})  T_{Y-y}(x_{ab},x_{a_0a_1})
T_{Y-y}(x_{ab},x_{b_0b_1}),
\end{align} which may be a useful form to include effects beyond
the BFKL evolution.

Finally, we consider how the approach to saturation is modified in
the presence of power--law correlations in the target.
 The scattering amplitude of two dipoles off a large onium of size $x_{01}$ at
 small impact parameter can be computed similarly
as before \begin{align}& T_Y^{(2)}(x_{01};x_{a_0a_1},x_{b_0b_1})
\sim
 \alpha_s^4 x_{a_0a_1}^2x_{b_0b_1}^2n_Y(x_{01};x_{a_0a_1},x_{b_0b_1})  \nonumber \\
 & \qquad \qquad \qquad \qquad \sim T_Y(x_{01},x_{a_0a_1},b\approx 0)
 T_Y(x_{01},x_{b_0b_1},b\approx 0)
 \left(\frac{x_{01}}{x_{ab}}\right)^{2(2\gamma_a-\gamma)}.
 \label{kkk} \end{align}
 The power law correlation Eq.~(\ref{kkk}) is remarkable in view of the fact that
  the single dipole distribution has essentially no $b$--dependence
 deep inside the dipole $x_{01}$. Due to the enhancement factor,
 $\left(\frac{x_{01}}{x_{ab}}\right)^{2(2\gamma_a-\gamma)} \gg 1$, the problem
  of unitarity for $T^{(2)}$ is severer than that for $T$.  The condition
  $T^{(2)}\le 1$ is roughly equivalent to requiring that the
  exponential factor of $n^{(2)}$ vanishes along the saturation line
  $ x_{a_0a_1}=x_{b_0b_1}=1/Q_{pairsat}$ \begin{align} 2\chi(\gamma_a)Y-\gamma \ln
  \frac{x_{01}^2}{x_{ab}^2}-2\gamma_a \ln
  (x_{ab}^2 Q_{pairsat}^2)=0.
  \end{align} Solving this equation with the conditions
  Eqs.~(\ref{15})-(\ref{17}), one finds that  \begin{align}
 Q^2_{pairsat}=\frac{1}{x_{01}^2}e^{\frac{\chi(\gamma_a)}{\gamma_a}Y}
  \left(\frac{x_{01}^2}{x_{ab}^2}\right)^{1-\frac{\gamma}{2\gamma_a}}
  =\frac{1}{x_{ab}^2}\exp
  \left(\frac{ \chi'(\gamma)-\frac{\chi(\gamma)}{\gamma}}
  {\frac{\chi'(\gamma)}{\chi'(\gamma_a)}
  -\frac{2\gamma_a}{\gamma}}Y\right). \label{last} \end{align} Since $\gamma$ and $\gamma_a$ depend
   on coordinates, there is not a unique way of writing $Q_{pairsat}$. The first
  expression shows that,  when $\gamma_a \approx \gamma_s$,
  the onset of unitarity corrections is much earlier than the
  single dipole scattering case, $Q_{pairsat}\gg Q_s$ (c.f.,  Eq.~(\ref{qsat})).
   This is quite a contrast to the result  $Q_s\approx Q_{pairsat}$
   which would follow from the
   assumption of factorization $T^{(2)}\approx T^2$.
  The second expression of Eq.~(\ref{last}) emphasizes\footnote{The
  factor in the exponential
  can be shown to be positive for $\gamma_a\ge \gamma_s$.} that $Q_{pairsat}$ is not an
  intrinsic quantity of the target, but
  depends rather sensitively on the configuration of the projectile.

\section*{Acknowledgments}
 We are grateful to Edmond Iancu and Gregory Soyez for
 many discussions. Y.~H. thanks  Lev Lipatov for helpful conversations.
 We thank the Galileo Galilei Institute for
 Theoretical Physics for the hospitality and the INFN for partial
 support during the completion of this work. The work of A.~M. is
 supported in part by the U.S. Department of Energy.

\end{document}